# Digital Data Security and Copyright Protection Using Cellular Automata

[1]M.Khlaif, [2]M.Talb

[1, 2] Academy of Postgraduate Studies

Benghazi University Benghazi, Libya

**Abstract**
The emergence of many challenges and the rapid development of the means of communications and computer networks and the Internet. Digital information revolution has affected a lot on human societies. Data today has become available in digital format (text, image, audio, and video), which led to the emergence of many opportunities for creativity for innovation as well as the emergence of a new kind of challenges.

## 1. Introduction

Computer networks vulnerable to penetration and thus steal or modify digital data. For this we need the security of digital data and the means to protect the property rights of these data led to more interest in the development of new methods for watermark signs. Our goal in this research is to solve the problem of data security and protection of property rights of digital data by designing a new model of complex systems to mix digital data and watermark. We will examine the cellular machinery and model of different potentials to solve this problem. Digital data is any data in digital form any data storing and handling by modern digital devices such as computer or mobile phone. And these data may represent texts such as books and, magazines as they represent the voice data such as songs and the like may represent images or video data. Or may represent computer programs.

The ease of copying and transmission of digital data leads to the need for ways to protect the property rights of these data are found. Today many ways to protect the property rights of the data and there are several commercial programs that do so. And protect the rights of ownership of digital data by hiding information about the original owner of the data within this data (ie signing or company logo, for example). One of these concealment methods called watermark) within its own digital data so that it can be using a certain algorithm to obtain this hidden data later to prove ownership of the owner. There are two types of watermark, the first type visible and can be seen, Figure 1,and the second type is hidden and cannot be seen. The first type is used in special cases where it can remove the watermark by using processes and image processing watermark for not knowing their place. Must be fixed watermark cannot be removed; it must be resistant to various types of attack, which may be from people such as data hackers. Must resist the watermark various processes common to image processing such as pressure and filtration and Engineering modifications. There are many applications for water mark. For example, television broadcasting, electronic books, electronic fingerprinting and so on.

1.1 Research Problem

With the rapid development of communications and computer networks and the Internet and the information revolution, digital data are available today in digital format (text, image, voice, and video) and this leads to the emergence of many opportunities for innovation and creativity as well as the emergence of challenges of a new type. Computer networks vulnerable to penetration and thus steal or modify these digital data and this leads to the need to protect digital data and the means to protect the property rights of the owners of these data. All this leads to more interest in the development of new methods for water signs. Our goal in this research is to solve the problem of data security and protection of property rights of digital data by designing a new model of complex systems to mix digital data and work watermark. We will examine the cellular machinery model and potential to solve this problem.

In this paper, we will focus our attention on the hidden watermark that cannot be seen or know their place, but by the owner of the data by using a special algorithm water mark. In this research, we will develop an algorithm of determination using the cellular machinery to form work watermark efficient and effective for different kinds of digital data (text, audio, image, video).

1.2 B. History

Data mean to hide here hide process digital data, among other digital data is called the cover or container. This means making the data that has been hidden (not visible), but anyone who knows the algorithm to show hidden data [1]. The process can be used to hide secret data to communicate between the parties so that the party will send the message to hide the message within the data and send it and the other side is the receipt of data and extract the hidden message using a specific algorithm to be agreed upon between the two parties.

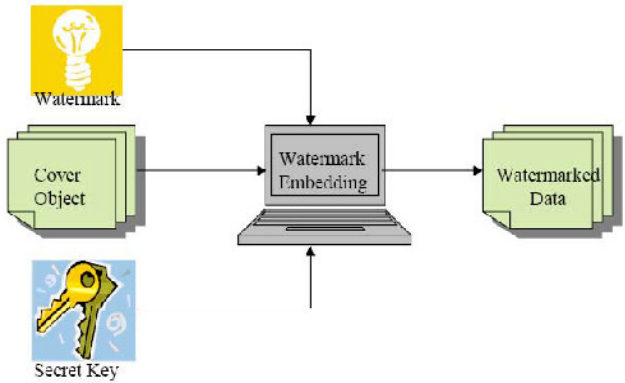

Fig 1 Embedding Watermark [9]

The watermark is one of the most important applications to hide data which is about changes in the original work (digital data) to add owner information to work (digital data as well) within this work (Fig. 1). Different watermarking method camouflage stenography in camouflage hidden data while interested in the watermark interest is the protection of the work itself and not hidden data [2].

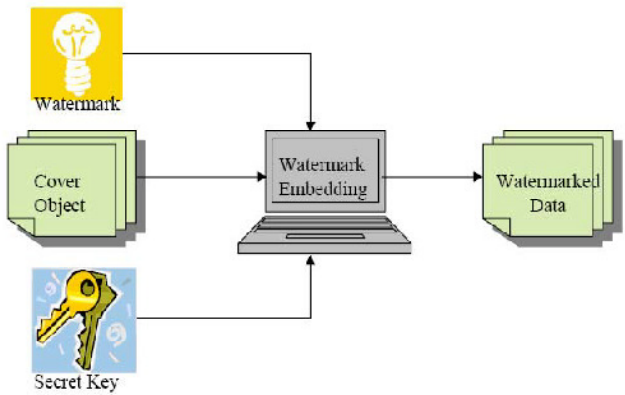

Fig 2 Extracting Watermark [9]

Watermark must resist any amendment to the original work. May be exposed watermark for many different threats that attempt to remove or modify the watermark. Including image compression, start jamming, conversion to another domain, different filters, sports transfers, tag, pseudo revoke parts of the image randomly had lost a large part of the watermark can increase the effectiveness of the watermark by repetition, temporal or spatial fingerprints, and use more than one algorithm at the same time [2].

## 2. Cellular machinery

Cellular machinery is a simple dynamic model of computing [3,4,5]. This model gives the behavior and complex output and interesting. This model can imagine a group of computers with similar special purpose and containing the same program and has the memory bear primary values change from one issue to another. These computers interact with each other through neighborly relationship vary from one issue to another (see Table4).There are many algorithms in the watermark. To space constraints we will not detail the traditional algorithms, where it was covered adequately in many references, so we will focus on new business related. Of traditional algorithms algorithm to decide the least important Least Significant Bit an algorithm working to conceal watermark data within the least significant bits of the original data [6]. Algorithms are also known algorithm to convert the cosine of the angle Cosine Transform, where this algorithm converts data from its original space into a new area by using the equation of the cosine of an angle [7].

Table 1. Primary Values

| X-1 | X | X+1 |
|---|---|---|

The random mixing and Shuffling a simple and effective algorithms in the process of masking data, the algorithm re-arrangement places the data in a certain way and thus hide the original data could not be obtained except by way of changing the order in [8] The use of cellular machinery in information hiding and watermark of new ideas in this thread [9,10,11,12]. Our work has been compared to the work of researches in [11,12] have been selected this work because it new and important . Previous research of this topic depends on the applications of the watermark in pictures. The idea is the work cell machines model the size of the original image and are filled random values (here must save function values start random values seeds) changed at every stage generation according to certain cellular machinery base rule.

At each stage is moved elements of the image pixels that the value of the corresponding sites in the sample cell machinery equal to one and put in a new order. The final product is a new picture quite different from the original image, but have the same data values (both of the two images of the same form statistical data histogram). After that, add watermark image through a logical collection (OR operation) of the values of the picture elements corresponding certain proportions. After that algorithm is reversed. Machines cell to retrieve the original image and thus scattered elements image watermark across the image original sort of jamming light. To restore the watermark must reverse the previous steps with the knowledge of a number of parameters such as values function start random values seeds and base machine cell as well as the number of stages. The advantages of this method are its simplicity and use of cellular machines and can therefore work computer hardware and thus speed in the application. Disadvantages of this method used for two-dimensional cellular machinery and thereby increase the time complexity of the algorithm Time complexity. Defects also increased jamming the greater the size of the watermark.

## 3. Research Idea

**Table 2 Neighborly Relationship**

| 7 | 6 | 5 | 4 | 3 | 2 | 1 | 0 | state |
|---|---|---|---|---|---|---|---|---|
| 111 | 110 | 101 | 100 | 011 | 010 | 001 | 00 | Old state |
| 101 | 100 | 101 | 100 | 001 | 010 | 011 | 010 | New state |

Purpose of this paper is by solving the problem of data security and protection of property rights of digital data by designing a new model of complex systems to mix digital data and work watermark. In this work we will examine the cellular machinery model and its potential to solve this problem will be our focus and attention on the hidden watermark that cannot be seen or know their place only by the owner of the data by using a special algorithm water mark. In this research, we will develop a new algorithm of determination using Model cellular machinery to work the hidden watermark.

## 4. Algorithms

Table 3. Algorithm for Converting Cosine

```
Algorithm : Scramble

 Input:   I = index of the image(height, width or depth).

 Output: J= the scrambled indices
         S=the seed;
Create three one dimensional arrays with length I:
array P contains the index numbers; array A
random binary array;
E array of ones.
n=1; //new index

for 20 iterations //good enough
  for each elements in P
    Apply Cellular Automata rule #7
    if A(current)==1 and E(current) ~=0
     J(n)=P(current)    ;
     E(current)= 0;
     n=n+1;
    if A==J // no new updates
     Break;
```

## 5. Results

We apply our idea by using environment MATLAB [13] that offers many possibilities and tools ready for image processing. After doing many experiments on a different set of pictures (bilateral, gray and colored) made sure the different data in the image quality details and different sizes. The use of cellular machinery with Rule number (7) a variable number of stages have been applied to several types of attack on the watermark to test durability. Types of attack that have been applied: - interference by up to 40% and random deletions of the image in addition to the way JPEG image compression rates ranging from 30% to 80%. We also re-tests several times to make sure of the validity of the results. In all our experiments the results were better by more than 4% from the previous work, were comparable for By the difference between the original image and the resulting image after the mixing process by the difference between the values of grayscale two, according to the following equations:

$$GD_{i,j} = \frac{1}{4} \sum_{i',j'} (P_{i,j} - P_{i',j'})^2 \quad (1)$$

$E(GD_{ij}) = \sum_{i=2}^{M-1} \sum_{j=2}^{N-1} GD_{ij} / (M-2)(N-2)$ ( 2)

Where equations applied to P-dimensional image M and N. GD is the difference in the values of grayscale Gray Difference between the current image element and adjacent picture elements. E is the average GD all elements except the parties. GDD is the difference between the average differences of the original image and the image resulting from the mixing process ..Can refer to research [11] for more details on these equations. Form (4) gives an overview of the results obtained and the agenda (1) gives a comparison between the results of our search and search results [11] which is applied Rule 224, and the search results [12], which applies the base for the Game of Life. Also noticed that our best results in terms of the watermark can be seen by the human eye clearly even after the application of the various different types of attack designed to remove the watermark. Form is divided into two main columns. Each column contains the results of a picture and a different watermark (image Jean camera and the image of a man). In each column, the left part is a kind of attack and the right side contains the watermark extracted.

A. Original image, the watermark.

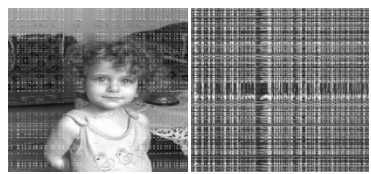

B. Original image after mixing - original

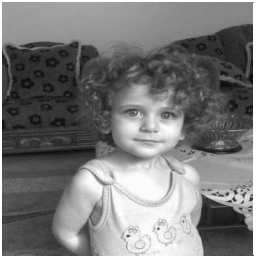

C. Image with the watermark.

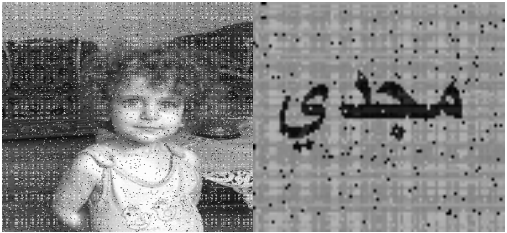

D. Original image watermark with jamming attack - the watermark extracted.

E. Original image watermark with random attack deleted - watermark extracted.

F. Original image watermark with attack JPEG compression – watermark extracted. Table (I) Comparison GDD values of the proposed model and previous models.

## 6. Conclusion and future work

In this research, we present a new type of model one-dimensional cellular machinery is characterized by simple and efficient and can be applied in the protection of property rights and security of digital data to computer networks. Proposed model has been applied to digital images as an example of digital data with the observation that method can be applied to any kind of data easily by converting data into one-dimensional matrix. Been confirmed by many scientific experiments that watermark hidden in our own way can withstand many different kinds of attack designed to remove such as noise and jamming deletion cropping and pressure compression. Experience has also shown that our model best by more than 4% from the previous business. Future works include more study of various cellular models of machines and increase the allocation of this model to suit the various types of digital data more efficiently. In addition to test this model and continue to improve the resistance to other types of different types of attack.